\documentclass[showpacs,prl,onecolumn,aps,superscriptaddress,preprintnumbers,nofootinbib]{revtex4}
\usepackage{amsmath,amssymb}
\usepackage{epsfig}
\usepackage{graphicx}
\usepackage{amsmath}
\usepackage{amsfonts}
\usepackage{epstopdf}
\def\slashchar#1{\setbox0=\hbox{$#1$}     		
   \dimen0=\wd0                                 	
   \setbox1=\hbox{/} \dimen1=\wd1               	
   \ifdim\dimen0>\dimen1                        	
      \rlap{\hbox to \dimen0{\hfil/\hfil}}      	
      #1                                        	
   \else                                        	
      \rlap{\hbox to \dimen1{\hfil$#1$\hfil}}   	
      /                                         	
   \fi}

\renewcommand{\vec}{\boldsymbol}
\newcommand{\beq}{\begin{equation}}
\newcommand{\eeq}{\end{equation}}
\newcommand{\bea}{\begin{eqnarray}}
\newcommand{\eea}{\end{eqnarray}}
\newcommand{\ba}{\begin{array}}
\newcommand{\ea}{\end{array}}

\def\eq#1{{Eq.~(\ref{#1})}}
\def\fig#1{{Fig.~\ref{#1}}}
\newcommand{\bas}{\bar{\alpha}_S}

\newcommand{\nn}{\nonumber}

\newcommand{\h}{\frac{1}{2}}

\newcommand{\Lb}{\left(}
\newcommand{\Rb}{\right)}
\def\pom{{I\!\!P}}

\begin{document}

\title{ Bose-Einstein correlations and $\mathbf{v_{2n}}$ and  $\mathbf{v_{2n-1}}$  in  hadron and nucleus collisions}
\author{E. ~Gotsman}
\email{gotsman@post.tau.ac.il}
\affiliation{Department of Particle Physics, School of Physics and Astronomy,
Raymond and Beverly Sackler
 Faculty of Exact Science, Tel Aviv University, Tel Aviv, 69978, Israel}
 \author{ E.~ Levin}
\email{leving@post.tau.ac.il, eugeny.levin@usm.cl}
\affiliation{Department of Particle Physics, School of Physics and Astronomy,
Raymond and Beverly Sackler
 Faculty of Exact Science, Tel Aviv University, Tel Aviv, 69978, Israel}
 \affiliation{Departemento de F\'isica, Universidad T\'ecnica Federico
 Santa Mar\'ia, and Centro Cient\'ifico-\\
Tecnol\'ogico de Valpara\'iso, Avda. Espana 1680, Casilla 110-V,
 Valpara\'iso, Chile} 
 \author{  U.~ Maor}
\email{maor@post.tau.ac.il}
\affiliation{Department of Particle Physics, School of Physics and Astronomy,
Raymond and Beverly Sackler
 Faculty of Exact Science, Tel Aviv University, Tel Aviv, 69978, Israel}
\date{\today}


\begin{abstract}
We show that Bose-Einstein correlations of identical particles in
 hadron and nucleus high energy collisions, lead to  long range
 rapidity correlations in the azimuthal angle. These correlations are
 inherent features of the CGC/saturation approach, however,
 their  origin is more general  than this approach. In framework of the proposed
 technique both even and odd $v_n$ occur naturally, independent
 of the type of target and projectile. We are of the opinion that it is premature to conclude that 
the appearance of  azimuthal correlations
  are due to the hydrodynamical behaviour of the quark-gluon plasma.
 \end{abstract}
 
 \preprint{TAUP-3007/16}

\maketitle

 One of the most intriguing experimental observations made at the LHC
 and RHIC, is the occurrence of the same pattern of  azimuthal angle correlations in
the  three types of  interactions: hadron-hadron, hadron-nucleus and
 nucleus-nucleus collisions. In all three reactions,  correlations
in the events with large density of produced particles,
 are observed between two charged hadrons, which are separated by 
 large values of rapidity  \cite{CMSPP,STARAA,PHOBOSAA,STARAA1,CMSPA,CMSAA,
ALICE}  and  these correlations do not depend on the rapidity
 separation of the particles. Due to
 causality arguments\cite{CAUSALITY}, two hadrons with large difference
 in rapidity between them, could only correlate  at  the early stage 
 of the collision and, therefore,  we expect that the correlations
 between two particles with large rapidity difference (at least the
 correlations in rapidity) are due to the partonic state with large
 parton density.  The CGC/saturation approach (see \cite{KOLEB}
 for a review) appears to be a natural candidate for the description of 
these
 correlations, as these correlations are strong in
 the dense colliding systems.
  However, unlike the large rapidity correlations,  the
 azimuthal angle correlations  can  originate from the collective
 flow in the final state \cite{FINSTATE}.  At first sight, 
 this source appears even more plausible, since $v_n$ with odd
 $n$ do not appear in the CGC/saturation approach.

\begin{figure}[ht]
    \centering
  \leavevmode
      \includegraphics[width=12cm]{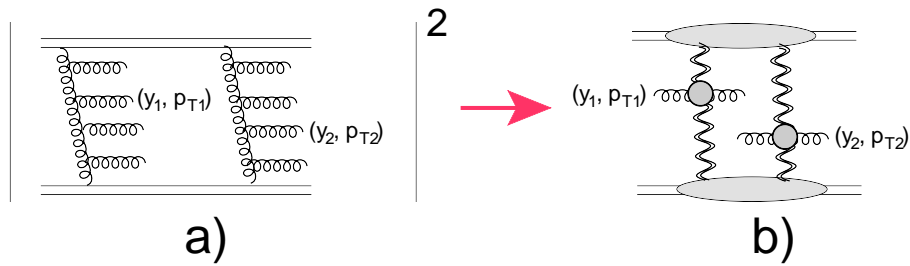}  
      \caption{ Production of two gluons with $(y_1,\vec{p}_{T1})$ and $(y_2,\vec{p}_{T2})$ in
 two parton showers (\protect\fig{2sh}-a).   \protect\fig{2sh}-b shows  the double inclusive
 cross section in Mueller diagram technique \protect\cite{MUDI}. The wavy lines denote the BFKL Pomerons}
\label{2sh} 
   \end{figure}

\begin{figure}[ht]
    \centering
  \leavevmode
      \includegraphics[width=12cm]{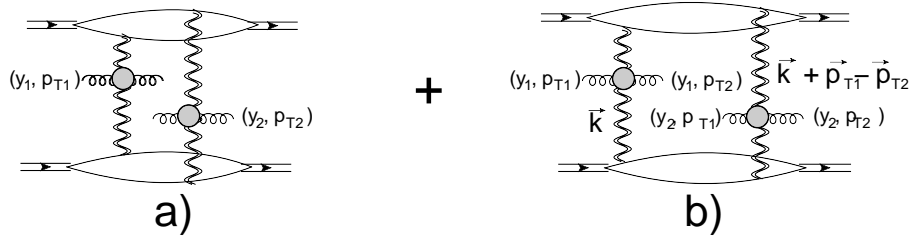}  
      \caption{ Production of two identical  gluons with $(y_1,\vec{p}_{T1})$
 and $(y_2,\vec{p}_{T2})$ in two parton showers.  The diagrams in the
 Mueller diagram technique
 \protect\cite{MUDI} are shown in \fig{2shiden}-a and \fig{2shiden}-b.
 The wavy lines denote the BFKL Pomerons\cite{BFKL,LI}.} 
      
      \label{2shiden}
       \end{figure}

In this article we show that the long range rapidity correlations
 in the azimuthal angle, arise naturally  from the Bose-Einstein
 correlations of produced identical particle in high energy collisions.
 They originate from the initial state wave function of the colliding particles,
 and they are   features characteristic  of the CGC/saturation 
approach. 
However,
 their occurrence is  more general,  and can be estimated in other frameworks.
 The
 long range rapidity correlations stem from the production of two parton
 showers in QCD (see  \fig{2sh}).  The structure of the parton shower in QCD
 is described by the exchange of the BFKL Pomeron,  while the upper and low
 blobs in \fig{2sh}-b require  modeling, due to our poor theoretical
 knowledge of the confinement of quarks and gluons.  However, if two
 produced gluons have the same quantum numbers, we need to take into
 account an  additional Mueller  diagram\cite{MUDI} of \fig{2shiden}-b
 in which two gluons with
 $(y_1,\vec{p}_{T2})$ and  $(y_2,\vec{p}_{T1})$ are produced. 
 When $\vec{p}_{T1} \,\to\,\vec{p}_{T2}$ the two  production processes 
become
 identical, leading to the cross section $\sigma\Lb \mbox{two identical
 gluons}\Rb = 2 \sigma\Lb \mbox{two different  gluons}\Rb$, as one expects.
 However, when $|\vec{p}_{T2} -   \vec{p}_{T1}| \gg 1/R$ where $R$ is the
 size of the emitter, the interference diagram becomes small and can be
 neglected.
   
 At first sight the Mueller diagram of \fig{2shiden}-b in general case $y_1 \neq y_2$ and $\vec{p}_{T1} \neq \vec{p}_{T2}$   looks problematic,  since it describes  the interference between two different final states.  However, in the leading log(1/x) approximation(LLA) of perturbative QCD $\Lb\bas \Lb Y - y_1\Rb\Rb^{n_1}, \Lb\bas \Lb y_1 - 0\Rb\Rb^{n_2} ,\Lb\bas \Lb Y - y_2\Rb\Rb^{n_3} $ and $\Lb\bas \Lb  y_2 - 0\Rb\Rb^{n_4} $ contributions stem from the integration of the phase space  which for the gluon $(i)$ has the following form: $ d y_1 d^2 p_{Ti}$. In LLA we have the following ordering (for two parton showers production):
 \bea \label{LLA1}
 \mbox{first parton shower} & \rightarrow & Y\,>\,\dots \,> \,y_i\,>\,\dots\,>\,y_{n_1} \,>\,y_1\,>\,y_{n_2}\,>\,\dots \,> \,y_i\,>\,\dots\,>0;\nn\\
   \mbox{second parton shower} & \rightarrow& Y\,>\,\dots \,> \,y_i\,>\,\dots\,>\,y_{n_3} \,>\,y_2\,>\,y_{n_4}\,>\,\dots \,> \,y_i\,>\,\dots\,>0;
   \eea
 Integrating over $y_i$ and neglecting $y_i$ dependence of the production amplitude \cite{BFKL,LI} we obtain the contribution 
 \bea \label{LLA2}
    \frac{d \sigma^{\rm different \,gluons}}{d y_1 d^2 p_{T1}\,d y_2\,d^2 p_{T2}}\,\,&=&\,\,\sum^\infty_{n_1+n_2-2>2}\,\, \sum^\infty_{n_3+n_4-2>2}  \,\int d\Phi^{(1)}_{n_1 + n_2} d \Phi^{(2)}_{n_3+n_4}\,|A^{\rm different \,gluons}\Lb \{y_i,p_{Ti}\}; y_1,p_{T1}; y_2,p_{T2}\Rb|^2 \nn \\
       & = &\sum^\infty_{n_1+n_2-2>2}\,\, \sum^\infty_{n_3+n_4-2>2}\underbrace{\frac{\Lb\bas \Lb Y - y_1\Rb\Rb^{n_1}}{n_1!}\,\frac{\Lb\bas \Lb y_1 - 0\Rb\Rb^{n_2}}{n_2!} \,\frac{\Lb\bas \Lb Y - y_2\Rb\Rb^{n_3}}{n_3!}  \,\frac{\Lb\bas \Lb  y_2 - 0\Rb\Rb^{n_4} }{n_4!} }_{\rm integral\,\,over\,\,lomitudinal\,\,phase\,\,space} \nn\\
    &\times& \int \prod_i d^2p_{Ti}\, \,|A^{\rm different\, gluons}\Lb \{y_i=0,p_{Ti}\}; y_1=0,p_{T1}; y_2=0,p_{T2}\Rb\|^2 
    \eea
     where $d \Phi^{(1)}_{n_1+n_2}$ and $d \Phi^{(2)}_{n_3+n_4}$
 denote the phase space of the produced gluons in the first and second 
parton showers. 

    \eq{LLA2} represents the factorization of the longitudinal and
 transverse degrees of freedom, which is the principle characteristic
 of the LLA\cite{BFKL,LI}.
    
    For two parton showers the production amplitude $A^{\rm different\,
 gluons}\Lb \{y_i=0,p_{Ti}\}; y_1=0,p_{T1}; y_2=0,p_{T2}\Rb\,\propto\, 
   A^{(1)}_{n_1 n_2}\Lb \{y_i=0,p_{Ti}\}; y_1=0, p_{T1}\Rb \,  
 A^{(2)}_{n_3 n_4}\Lb \{y_i=0,p_{Ti}\}; y_2=0,p_{T2}\Rb $. Summing
 over $n_i$, we  obtain the Mueller diagram of \fig{2sh}-b.
    
      For identical particles 
    we need to replace
    \bea \label{LLA3}
 &&   A^{\rm different\, gluons}\Lb \{y_i=0,p_{Ti}\}; y_1=0,p_{T1}; y_2=0,p_{T2}\Rb\,\propto\nn\\
 &&\, A^{(1)}_{n_1 n_2}\Lb \{y_i=0,p_{Ti}\}; y_1=0,p_{T1}\Rb \,   A^{(2)}_{n_3 n_4}\Lb \{y_i=0,p_{Ti}\}; y_2 =0, p_{T2}\Rb \,\,\longrightarrow  \\
 &&  A^{\rm identical \, gluons}\Lb \{y_i=0,p_{Ti}\}; y_1=0,p_{T1}; y_2=0,p_{T2}\Rb\,=\,\nn\\
 &&  A^{\rm different\, gluons}\Lb \{y_i=0,p_{Ti}\}; y_1=0,p_{T1}; y_2=0,p_{T2}\Rb\,\,+\,\,  A^{\rm different\, gluons}\Lb \{y_i=0,p_{Ti}\}; y_2=0,p_{T2}; y_1=0,p_{T1}\Rb \propto \nn\\
 && A^{(1)}_{n_1 n_2}\Lb \{0,p_{Ti}\}; 0, p_{T1}\Rb \,   A^{(2)}_{n_3 n_4}\Lb \{0,p_{Ti}\}; 0,p_{T2}\Rb\,\,+\,\,     A^{(1)}_{n_1 n_2}\Lb \{0,p_{Ti}\}; 0,p_{T2}\Rb \,   A^{(2)}_{n_3 n_4}\Lb \{0 ,p_{Ti}\}; 0,p_{T1}\Rb  \nn
 \eea  
 
 We note that  in \eq{LLA3} we use the Bose-Einstein
 symmetry for the amplitudes which depend {\it only} on the transverse
 momenta of produced particles.  Summing over $n_i$, we obtain the
 contributions which are shown in \fig{2shiden}-a and   \fig{2shiden}-b.
 
   Estimates in the LLA,  which we discussed above,  are performed in 
the kinematic region where
 $\bas\Lb y_k - y_m\Rb\,\geq 1$ ($y_k  (y_m) = y_0 ,y_1, y_2,
 y_4=Y$ with $y_0 = 0$ ) including $\bas\Lb y_1 - y_2\Rb\,\geq 1$).
 Note that the calculations of Ref.\cite{KOLUCOR},  were done for 
  $\bas\Lb y_1 - y_2\Rb  \,\leq \,1$.

\begin{figure}[ht]
    \centering
  \leavevmode
      \includegraphics[width=16cm]{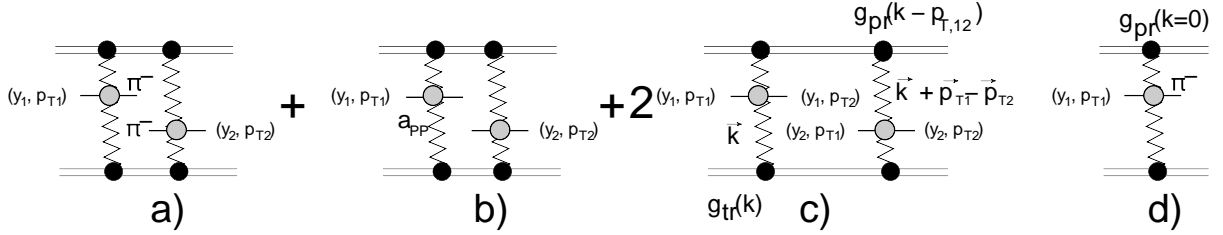}  
      \caption{The Mueller diagrams for  production of two identical  $\pi^-$ with $(y_1,\vec{p}_{T1})$
 and $(y_2,\vec{p}_{T2})$ in two parton showers (see \fig{2shpion}-a -
 \fig{2shpion}-c).  The diagram of \fig{2shpion}-d describes the inclusive production of pions. The zigzag  lines denote the soft Pomerons.} 
       \label{2shpion}
       \end{figure}

 The angular correlation stems from the diagram of \fig{2shiden}-b in 
which
 the upper BFKL Pomerons carry momentum
 $\vec{k} - \vec{p}_{T,12}$ with $\vec{p}_{T, 12}\,=\,\vec{p}_{T1}\,-\,
\vec{p}_{T2}$, while the lower BFKL   Pomerons have  momenta $\vec{k}$.
  In this article we demonstrate a mechanism for the appearance of these
 angular correlations in the framework of a simple approach: the soft
 Pomeron calculus\footnote{ The correlation of identical
 particles was investigated in the framework of the soft
 Pomeron calculus and the mechanism of the azimuthal angle correlation  that we discuss here,  has been proposed  in Ref.\cite{PION} for hadron and nucleus interactions.  Recently, it has been re-discovered in Ref.\cite{KOLUCOR} in the framework of the CGC approach.  We re-visit this formalism
 for calculations  of $v_n$  for odd and even $n$.}.
 The Mueller diagrams for the correlation between two
 $\pi^-$ are shown in \fig{2shpion}. As can be seen from
 \fig{2shpion} we use the eikonal model for  estimates
 of the amplitude for two soft Pomeron production. In the
 case of the nucleus target and/or projectile, this model
 corresponds to the Glauber model. 



 For the vertices of the soft Pomeron interaction with the projectile and target
 we use the following parametrizations:
 \beq \label{G}
 g_{pr}\Lb k^2\Rb \,=\,g^0_{pr}\,e^{- \h B_{pr}\,k^2_T};~~~~~~~~ g_{tr}\Lb k^2\Rb \,=
\,g^0_{tr}\,e^{- \h B_{tr}\,k^2_T}; 
 \eeq
 and for the vertex of pion emission from the Pomeron we use the simplest
 parametrization:
 \beq \label{A}
 a_{\pom \pom}\Lb p_{T1},p_{T2}\Rb\,=\, a^0_{\pom \pom}\, e^{- \h B_{e}\,\Lb 
 p^2_{T1}\,+\,p^2_{T2}\Rb}
 \eeq
  We have neglected the  possible dependence of $a_{\pom\pom}$ on $k^2$
 and $\vec{p}_{T, 12}$. 

 
 The contribution of \fig{2shpion}-c to the double inclusive cross section
 is equal to
 \bea \label{DI1}
 \frac{d \sigma}{d y_1 d^2 p_{T1}\,d y_2\,d^2 p_{T2}}\,\,&=&\,\,a^2_{\pom \pom}\Lb p_{T1},
 p_{T2}\Rb\,e^{2 \Delta_\pom Y}\,\frac{\Lb g^0_{pr}\Rb^2\,\Lb g^0_{tr}\Rb^2}{4 \pi^2}\int d^2
 k_T\,\exp\Big( - B_{tr} \,k^2_T\,- B_{pr}\Lb \vec{k}_T - \vec{p}_{T,12}\Rb^2\Big)\,\\
 &=&\,\Lb a^0_{\pom \pom}\Rb^2\frac{\Lb g^0_{pr}\Rb^2\,\Lb g^0_{tr}\Rb^2}{4 \pi\,\Lb B_{tr} +
 B_{pr}\Rb}\,e^{2 \Delta_\pom Y}\,\exp\Big( - \Lb  B_e  + B_R\Rb\,\Lb p^2_{T1}\,+\,p^2_{T2}\Rb\,-\,
2\,B_R\,p_{T1} p_{T2}\cos\Lb \varphi\Rb\Big)\nn
\vspace{0.2cm} 
  \eea
  where 
  \beq \label{BR}
  B_R\,\,=\,\,\frac{B_{pr}\,B_{tr}}{B_{pr}\,+\,B_{tr}}
  \eeq
  and $\Delta_\pom$ is the intercept of the soft Pomeron.
 The sum of all diagrams of \fig{2shpion} leads to
 \bea \label{DI2}
&&  \frac{d \sigma}{d y_1 d^2 p_{T1}\,d y_2\,d^2 p_{T2}}\,\,
 =\,\,\\
 &&\Lb a^0_{\pom \pom}\Rb^2\frac{\Lb g^0_{pr}\Rb^2\,\Lb g^0_{tr}\Rb^2}{4 \pi\,\Lb B_{tr}
 + B_{pr}\Rb}\,e^{2 \Delta_\pom Y}\,\exp\Big( -  B_e \,\Lb p^2_{T1}\,+\,p^2_{T2}\Rb\Big)
\,\Bigg\{ 1\,\,+\,\,\exp\Big(\,-\,B_R\,\Lb   p^2_{T1}\,\,-2\, p_{T1} p_{T2}\cos\Lb
 \varphi\Rb \,+\,p^2_{T2}\Rb\Big) \Bigg\}\nn
 \eea
 
       \begin{figure}[ht]
    \centering
  \leavevmode
      \includegraphics[width=12cm]{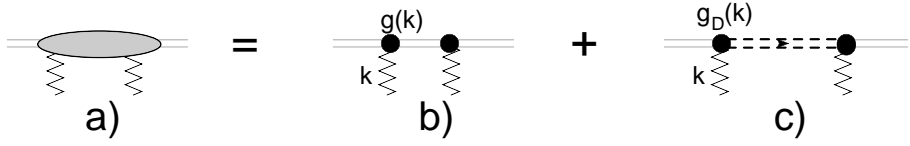}  
      \caption{ The structure of the  Pomeron-proton amplitude: 
\fig{strN}-b is the
 contribution of the eikonal approach,\fig{strN}-c is  
the diffraction dissociation contribution.} 
      
      \label{strN}
       \end{figure}


   The expansion of \eq{DI2}  contains all powers of $\cos\Lb 
\varphi\Rb$,
 or in other words, all $\cos\Lb n\, 
\varphi\Rb$ with even and odd $n$. The other feature of \eq{DI2} is that
 the double inclusive 
cross section does not depend on $y_1$ and $y_2$, displaying the long
 rapidity correlations
\footnote{Strictly speaking $B_R$ depends on $y_1$ and $y_2$ due to the
 shrinkage of the diffraction
 peak,  but we neglect this contribution since the slope of the 
Pomeron trajectory ($\alpha_\pom$)  is rather small.}

 We can re-write \eq{DI2}  in terms of the observables which can be 
measured:  the slopes of elastic scattering
 and the rapidity correlation function $C\Lb y_1, y_2\Rb$ defined as
 \beq \label{C}
 C\Lb y_1,y_2\Rb\,\,=\,\,\frac{1}{\sigma_{in}}\,\int d^2 p_{T1}\,d^2 p_{T2}\,
  \frac{d \sigma}{d y_1 d^2 p_{T1}\,d y_2\,d^2 p_{T2}} \Bigg{/}\Lb\frac{1}{\sigma_{in}}\,\int d^2 p_{T1}\,\frac{d \sigma}{d y_1 d^2 p_{T1}}  \Rb\,\Lb\frac{1}{\sigma_{in}}\,\int d^2 p_{T2}\,\frac{d \sigma}{d y_2 d^2 p_{T,2}}  \Rb
 \eeq
 It is more convenient to introduces  a correlation function $C\Lb y_1,p_{T1};
 y_2,p_{T2}\Rb$ as
 \bea \label{CPT}
&&C\Lb y_1,p_{T1}; y_2,p_{T2}\Rb\,=\,\,\frac{\frac{1}{\sigma_{in}} \frac{d
 \sigma}{d y_1 d^2 p_{T1}\,d y_2\,d^2 p_{T2}}}{\Lb\frac{1}{\sigma_{in}}\,\,
\frac{d \sigma}{d y_1 d^2 p_{T1}}  \Rb\,\Lb\frac{1}{\sigma_{in}}\,\frac{d \sigma}{d y_2 d^2 p_{T2}}  \Rb}\\
&&=\,C\Lb y_1,y_2\Rb\,\Bigg\{ 1\,\,+\,\,\frac{B_R}{B_R +B_e}\exp\Big(\,-\,B_R\,\Lb   p^2_{T1}\,\,-2\, p_{T1} p_{T2}\cos\Lb
 \varphi\Rb \,+\,p^2_{T2}\Rb\Big) \Bigg\} \nn
  \eea
  
  Using 
  \beq \label{II}
  \int^{2  \pi}_0 d \varphi\, e^{2\, p_{T1} p_{T2}\cos\Lb
 \varphi\Rb} \,\cos\Lb n\,\varphi \Rb \,=\,\,2\,\pi\,I_n\Lb 2\, p_{T1} p_{T2}\Rb,
 \eeq
 where $I_n\Lb z\Rb$ is the modified Bessel function of the first kind,
  we will decompose the term in $\{ \dots \}$ in  $C\Lb y_1,p_{T1};
 y_2,p_{T2}\Rb$  into Fourier modes in
 the relative azimuthal angle  $\varphi$ between two produced  pions:
  \bea \label{DVN}
&&  C\Lb y_1,p_{T1}; y_2,p_{T2}\Rb\,\propto\,1 \,\,+\,\,2\,\sum_{n=1}\,V_{n \Delta}\Lb p_{T1},p_{T2}\Rb\,
\cos\Lb n \varphi\Rb\nn  \\
&&  \mbox{with} ~~ V_{n \Delta}\Lb p_{T1},p_{T2}\Rb \,=\,   \frac{1}{2} I_n\Lb 2\,B_R\,p_{T1} \,p_{T2}\Rb
\frac{e^{-\,B_R\,\Lb  p^2_{T1} +  p^2_{T2}\Rb }}{ 1 \, +\,I_0 \Lb 2\, B_R \,p_{T1} \,p_{T2}\Rb e^{-\,B_R\,\Lb   p^2_{T1} +  p^2_{T2}\Rb}}    
   \eea 
   assuming that $B_e\, \ll\, B_R$.

   The coefficients $v_n\Lb p_T\Rb$ are equal to
   \beq \label{VN}
   v_n\Lb p_T\Rb\,=\,\frac{V_{n \Delta}\Lb p_{T},p^{\mbox{\footnotesize Ref}}_{T}\Rb }
{\sqrt{V_{n \Delta}\Lb p^{\mbox{\footnotesize Ref}}_{T},p^{\mbox{\footnotesize Ref}}_{T}\Rb}}\,=\,\frac{1}{\sqrt{2}}
 \frac{I_n\Lb 2\,B_R\,p_{T1} \,p^{\mbox{\footnotesize Ref}}_{T2}\Rb}{\sqrt{  I_n\Lb 2\,B_R\,\Lb \,p^{\mbox{\footnotesize Ref}}_{T2}\Rb^2\Rb}}\,\frac{\sqrt{1\, +\,I_0 \Lb 2\, B_R\, \Lb p^{\mbox{\footnotesize Ref}}_{T2}\Rb^2\Rb e^{-\,2\,B_R\,  \Lb p^{\mbox{\footnotesize Ref}}_{T2}\Rb^2}}}{ 1 \, +\,I_0 \Lb 2\, B_R \,p_{T1} \,p^{\mbox{\footnotesize Ref}}_{T2}\Rb e^{-\,B_R\,\Lb   p^2_{T1} +  \Lb p^{\mbox{\footnotesize Ref}}_{T2}\Rb^2\Rb}}    
    \eeq
   where the value of $p^{\mbox{\footnotesize Ref}}_{T}$ is determined by the experimental procedure.
 Fixing $p^{\mbox{\footnotesize Ref}}_{T} = p_T$ we obtain
   \beq \label{vN}
         v_n\Lb p_T\Rb\,=\,\frac{1}{\sqrt{2}} \,e^{- B_R \,p^2_T} \sqrt{  \frac{I_n\Lb 2\,B_R\,p^2_T\Rb}{1\, +\,I_0 \Lb 2\, B_R\,p^2_T\Rb e^{-\,2\,B_R\,p^2_T}}}
        \eeq
  \eq{vN} stems from the diagrams of \fig{2shpion}. However like-sign pion 
pairs contribute a third
 of total contribution to pion pair production. 
  This means that the double inclusive cross section is equal to
 \bea \label{DIX}
   \frac{d \sigma}{d y_1 d^2 p_{T1}\,d y_2\,d^2 p_{T2}}\,\,&=&\,\,   \frac{d \sigma}{d y_1 d^2 p_{T1}\,d y_2\,d^2 p_{T2}}\Lb \rm unlike\,pairs\Rb \,\,+\,\,   \frac{d \sigma}{d y_1 d^2 p_{T1}\,d y_2\,d^2 p_{T2}}\Lb\rm identical\,pairs\Rb\,\,\nn\\
   &=&\,\,\frac{d \sigma}{d y_1 d^2 p_{T1}\,d y_2\,d^2 p_{T2}}\Lb \rm unlike\,pairs\Rb\,\,\Big( 1\,+\,\frac{1}{3}    C\Lb p_{T1}; p_{T2}\Rb  \Big)
   \eea

    Therefore, \eq{vN} has to be multiplied by factor  $1/3$. 
  
  In \eq{BR} $B_{pr}$ and $B_{tr}$ can be expressed in terms of the slope 
for elastic cross section
 for projectile-projectile and target -target scattering, respectively:
 $B_{pr} = \h B^{\rm el}_{\rm pr\,-\,pr}$ and $B_{tr} = \h B^{\rm el}_{\rm tr\,-\,tr}$.
  
\begin{figure}[ht]
  \begin{tabular}{cc}
  \leavevmode
   \includegraphics[width=8cm]{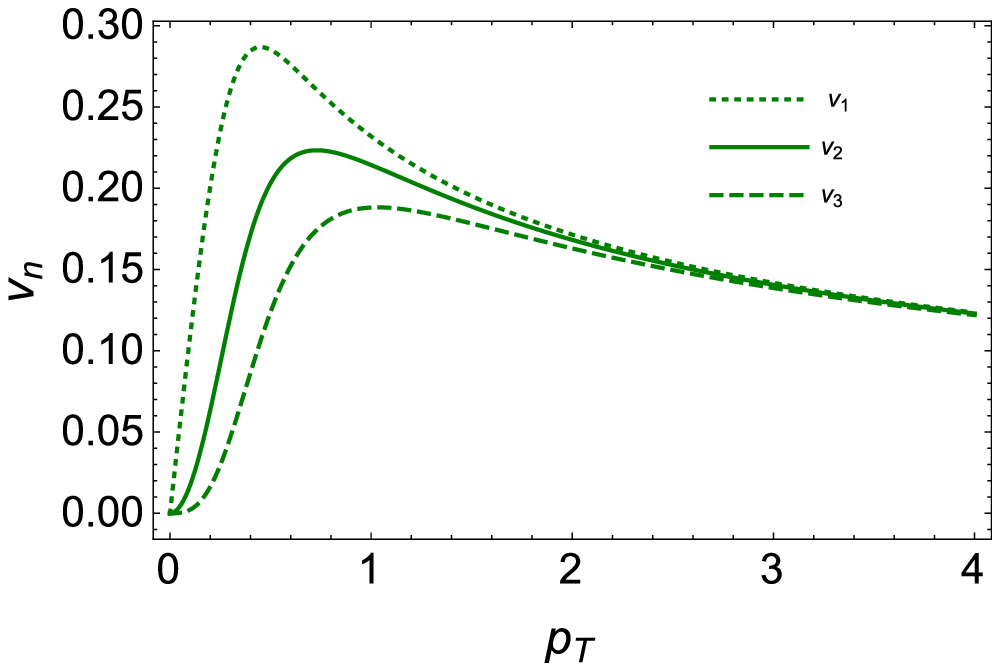}  & \includegraphics[width=8cm]{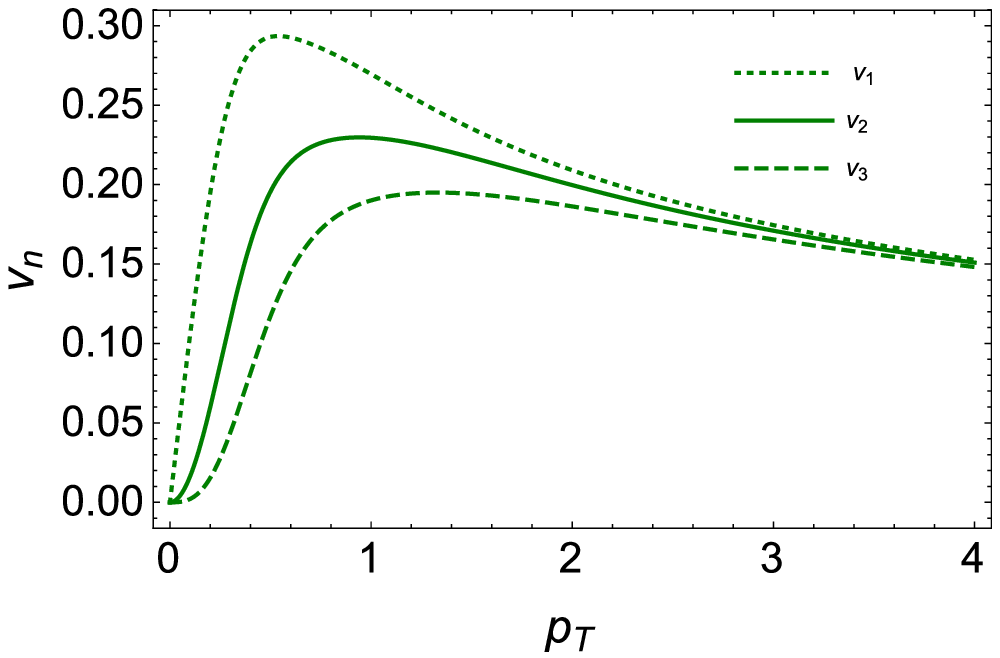} \\
   \fig{vn}-a &\fig{vn}-b\\
   \end{tabular}
   \caption{ $v_n$ versus $p_T$ using \protect\eq{vN}: \protect\fig{vn}-a using the eikonal structure of 
 Pomeron-hadron amplitude (see \protect\fig{strN}-b)) while  \protect\fig{vn}-b includes the process of
 diffraction dissociation (see \protect\fig{strN}-c).}
      \label{vn}
       \end{figure}

  For proton-proton scattering at W = 7\, GeV,       $B_{pr} = B_{tr} =  
\h B^{\rm el}_{\rm 
p\,-\,p}\,=\,10\,GeV^{-2}$ \cite{TOTEM}, which lead to $B_R\,=\,5\,GeV^{-2}$.
 Plugging this value in \eq{vN} we obtain the $v_n$ shown in \fig{vn}. One 
can 
see that we obtain
 sufficiently large values of $v_n$, which  are concentrated at rather large 
 values of $p_T$. The width of $p_T$ distribution will increase
 if we include a more complicated structure of the Pomeron-hadron amplitude
 (see \fig{strN}) and
 include  diffraction dissociation processes (see \fig{strN}-c),
   parametrizing $g_B(k) = g^0_D\exp\Lb - h B_D\,k^2\Rb$; \eq{vN} will
 have the following form:
 \beq \label{vND}
         v_n\Lb p_T\Rb\,=\, \frac{1}{3\,\sqrt{2}}\, \ \sqrt{\frac{I_n\Lb 2\,B_R\,p^2_T\Rb\,e^{- 2\,B_R \,p^2_T}\,\,+\,\,\frac{\sigma_{sd}\,B^{sd}_D}{\sigma_{el}\,B^{el}}\,I_n\Lb 2\,B_D\,p^2_T\Rb\,e^{- 2\,B_D \,p^2_T} }{1\, +\,I_0 \Lb 2\, B_R\,p^2_T\Rb e^{-\,2\,B_R\,p^2_T}\,\,+\,\,\frac{\sigma_{sd}\,B^{sd}_D}{\sigma_{el}\,B^{el}} I_0 \Lb 2\, B_D\,p^2_T\Rb e^{-\,2\,B_D\,p^2_T} }} 
         \eeq  
  where  $\sigma_{sd}$ denotes the cross section of the single
 diffractive production,
\footnote{For our estimates we took all cross sections from Ref.\cite{GLM2CH},}
 $B^{sd}_D$ the slope of the differential cross section
 for diffraction dissociation  is roughly  equal to $\h B^{el}$, and $B_D$ is  the slope
 of Pomeron-hadron vertex for diffraction dissociation (see \fig{strN}-c).
 The value of $B_D$ has been
 evaluated in Ref.\cite{KOTE}, and it is equal $\approx 1 \,GeV^{-2}$.
 \fig{vn}-b shows the calculation
 using \eq{vND} with $\frac{\sigma_{sd}\,B^{sd}_D}{\sigma_{el}\,B^{el}}$.
 One can see that the $p_T$
 distribution becomes broader. 

     
     In this article  our goal was not to describe the experimental data,
 but to demonstrate that a simple model leads to  reasonable  
values of $v_n$
 for proton-proton scattering. From the general expression of \eq{vN} we see
 that the estimates are independent of the type of projectile and target.
 We have not attempted to obtain an estimate for hadron-nucleus and
 nucleus-nucleus scattering, since the Gaussian approximation for
 $g(k^2)$ is not suitable for these reactions. Nevertheless,  in this 
oversimplified model,    
 \eq{BR} shows that for the hadron-nucleus interaction, the value
 and $p_T$ dependence of $v_n$ are determined by the size of
 proton, rather than the size of the nucleus.
  Realistic estimates from a model based on CGC/saturation approach,
 in which we successfully described the diffractive physics, as well as
 main features of the multiparticle production reaction\cite{
 GLM2CH,GLMNI,GLMINCL,GLMCOR} will follow in the near future.
  We wish  to point out that in CGC approach, pions
 originate  from the gluon jet decay, and we expect the same
 strength of correlations  both for like-like and unlike-like
 pion pairs as  is seen in experiments. To illustrate this
 it is enough to note that production of two like-like pairs
 of $\rho$-resonances,  that  dominate the inclusive production,
 say $\rho^0 \rho^0$ + $\rho^+ \rho^+$, generate the same numbers of
 $\pi^+\pi^+$ and $\pi^+ \pi^-$ pairs.
 
We proposed a mechanism for the long range rapidity azimuthal
 angle correlations which is general, simple and has a clear
 relation to  diffractive physics, unlike the hydrodynamic approach,
 which is suited to describe    only  processes of multiparticle
 generation.  We  demontsrated that this mechanism leads to the value
 of $v_n$ both for even and odd $n$, which are of the order of measured
 values for proton-proton collisions. We believe that it is  premature
 to conclude that the occurrence of  angular correlations is a strong
 argument in support of the hydrodynamical  behaviour of the quark-gluon 
plasma.

   {\bf Acknowledgements} 
  We thank our colleagues at Tel Aviv University and UTFSM for
 encouraging discussions. Our special thanks go to    
Carlos Cantreras, Alex Kovner and Michel  Lublinsky for
 elucidating discussions on the
 subject of this paper. This research was supported by the BSF grant   2012124, by    Proyecto Basal FB 0821(Chile) ,  Fondecyt (Chile) grant  1140842 and  by CONICYT grant PIA ACT140.

\end{document}